\documentstyle[amssymb,twocolumn,prl,aps,epsf]{revtex}

\begin{document}  

\twocolumn[\hsize\textwidth\columnwidth\hsize\csname
@twocolumnfalse\endcsname

\title{Metal-insulator transition in the double perovskites}
\author{A. A. Aligia, P. Petrone, J. O. Sofo, and B. Alascio}
\address{Comisi\'on Nacional de Energ{\'\i}a At\'omica,\\
Centro At\'{o}mico Bariloche and Instituto Balseiro, 8400 S.C. de Bariloche,%
\\
Argentina.}
\date{Received \today }
\maketitle

\begin{abstract}
We construct an effective Hamiltonian for the motion of electrons among the
transition metal ions of ordered double perovskites like Sr$_{2}$FeMoO$_{6}$%
, in which strong intra-atomic Coulomb repulsion $U$ is present in only one
of the inequivalent transition metal sites. Using a slave-boson formalism,
we construct a phase diagram which describes a charge transfer transition
between insulating and metallic behavior as the parameters of the model are
changed. The parameters for Sr$_{2}$FeMoO$_{6}$ are estimated from
first-principles calculations and a transition to the insulating state with
negative pressure is obtained.
\end{abstract}


]
\narrowtext

Two mechanisms contribute to the magnetoresistance (MR) of Mn Perovskites:
one of them is an intrinsic mechanism that results from the quenching of
spin scattering of the carriers by localized spins. It dominates the high
field MR and is most effective at temperatures of the order of the Curie
temperature $T_{c}$. The other mechanism which is most effective at low
fields, is due to the lowering of spin scattering at interphases between two
regions in the material that have different orientation of their
magnetization. This mechanism is also temperature dependent and is larger at
temperatures well bellow $T_{c}$, where the polarization of carriers is
large. This explains the interest on materials where $T_{c}$ is appreciably
larger than room temperature.

The report by Kobayashi {\it et al.}\cite{Kobayashi} that the double
perovskite Sr$_{2}$FeMoO$_{6}$ with $T_{c}$ of about 450 K is half metallic,
with an appreciable low field magnetoresistance has renewed interest in
these compounds. Furthermore, they open new questions concerning the nature
of the electronic structure.

In an ionic picture, assuming that all O ions are O$^{2-}$, one can imagine
the $3d$ transition metal ions as triply or doubly ionized, and Mo to be Mo$%
^{5+}(4d^{1},S=1/2)$ or Mo$^{6+}(4d^{0},S=0)$. In this picture Fe$%
^{3+}(3d^{5},S=5/2)$, would order antiferromagnetically with Mo to produce a
magnetization saturation of $4\mu _{B}$ as observed or Fe$^{2+}(3d^{6},S=2)$%
, could order ferromagnetically to produce the same magnetization. In the
extreme situation with no carriers on the Mo ion (Fe$^{2+}$ and Mo$^{6+}$),
one would expect that the system is insulating, because of the strong
intra-atomic Coulomb repulsion in Fe ($U\sim 7$ eV \cite{Boc}) compared with
the effective Mo-Fe hopping ($V\sim 0.39$ eV from our fits explained below
or $V\sim 0.25$ eV \cite{Millis}). Actually, there is a certain degree of
covalency between O and the transition metal ions \cite{Guille}. Applying
perturbation theory to this insulating state, the Mo valence becomes $%
v_{Mo}=6-24[t_{Mo-O}/(\epsilon _{Mo}-\epsilon _{O})]^{2}$, where $t_{tm-O}$
is the hopping between transition-metal $t_{2g}$ and O $p_{\pi }$ orbitals,%
\cite{exc} and $\epsilon _{X}$ is the on-site energy for atom $X$. The
magnetic moment of Mo $\mu _{Mo}$ remains zero in second order in $t_{Mo-O}$%
. This shows that $\mu _{Mo}$, rather than $v_{Mo}$ can indicate if the
system is insulating or near a metal-insulator transition. In fact, the
observed Mo-3d chemical shift is practically identical to that of MoO$_{3}$,
but $v_{Mo}<6$ \cite{Guille}. For Fe, $v_{Fe}-2=4-\mu
_{Fe}=8[t_{Fe-O}/(\epsilon _{Fe}-\epsilon _{O})]^{2}$.

While Sr$_{2}$FeMoO$_{6}$ is in fact metallic, neutron diffraction
experiments obtained magnetic moments consistent with an insulating state: $%
\mu _{Mo}=0\pm 0.1$ $\mu _{B}$ and $\mu _{Fe}=4\pm 0.1$ $\mu _{B}$,
indicating that the system is near a metal-insulator transition. Our band
structure calculations give a metallic state with $\mu _{Mo}\simeq 0.2$ $\mu
_{B}$, and with opposite sign as $\mu _{Fe}$. The fact that the experimental 
$\mu _{Mo}$ is lower, points out that the real system is nearer to the
insulating state than the predictions of these calculations. Furthermore,
substitution of Mo by Re seem to increase $T_{c}$ but renders the material
insulating. \cite{Gopa} . In addition, substitution of Fe by Co or Mn for
example makes the compounds antiferromagnetic and insulating \cite{Itoh}.
Also, the physical properties (conductivity, magnetization) are very
sensitive to sample preparation. Thus, it is of great interest to understand
when we can expect a metallic or insulating behavior in similar systems.

In this paper, we intend to address the general question of when one should
expect insulating or metallic behavior in the ordered double perovskites
structures.

In order to investigate the physical properties of these materials it is
necessary to gain knowledge in their intimate electronic structure,
including the effects of correlations. We build a tight binding Hamiltonian
to describe their electronic structure. This Hamiltonian is based on the
calculated energy bands, as explained later, and is reduced to the minimum
set of relevant parameters. Since the low-energy properties of the compounds
are determined by the bands crossing the Fermi energy, the only interesting
orbitals are the $t_{2g}$ orbitals in the interpenetrating simple cubic
lattices of Fe and Mo. Itinerant electrons tunnel from the three $t_{2g}$
orbitals of Fe to the same $t_{2g}$ orbitals of nearest neighboring Mo, and
between Mo and Mo with transfer energies $V$ and $\ V^{\prime }.$ In the
ferromagnetic phase, which will be studied here, only spin down electrons
can jump since the spin up orbitals at each Fe site are already filled. We
consider these orbitals to be frozen. The same approach has been followed in
a recent paper \cite{Millis}. The resulting effective Hamiltonian, which
contains strong correlations, is treated in a slave-boson approximation \cite
{kot,bal}.

\smallskip For our band structure calculations we use the Full-Potential
Linearized Augmented Plane Wave method (FP-LAPW) \cite{blaha95}. In brief,
this is an implementation of density functional theory using the generalized
gradient approximation (GGA) of the exchange and correlation potential.\cite
{perdew96} The Kohn-Sham equations are solved using a basis of linearized
augmented plane waves.\cite{singhbook} Local orbital extensions to the LAPW
basis\cite{singh91} are used to describe the 3$s$ and 3$p$ orbitals of Fe,
the 4$s$ orbitals of Mo and Sr. We use a well converged basis set of around
1260 plane waves and a sampling of the Brillouin zone (BZ) of 343 points,
corresponding to 20 in the irreducible wedge (IBZ). We use a muffin-tin
radius of 2.2 Bohr for Fe, 2.0 Bohr for Mo, 2.3 Bohr for Sr, Mo and 1.5 Bohr
for O.

To construct an effective Hamiltonian for the description of the motion of
electrons among transition-metal ions, we consider a simple cubic lattice
occupied by these ions, of lattice parameter $a$. This lattice consists of
two interpenetrating f.c.c. sublattices, one occupied by the Mo ions, and
the other by the Fe ions, in such a way that the nearest neighbors (NN) of
each Mo ion lie in the other sublattice and vice versa. This may be extended
to other transition metals. In between each two transition-metal ions lies
an O ion. The O degrees of freedom can be eliminated through a low-energy
reduction process (see for example, Ref. \cite{exc,sim2} and references
therein). If the difference between on-site energies $\epsilon
_{Mo}-\epsilon _{O}$ and $\epsilon _{Fe}-\epsilon _{O}$, is large compared
with the absolute value of $t_{Mo-O}$ and $t_{Fe-O}$, this can be done by
perturbation theory. The $xy$ orbitals acquire an effective hopping $V$ to
the $xy$ orbitals of the four NN in the $x-y$ plane due to a process of
second order in the hopping between $xy$ and $p_{\pi }$ orbitals \cite{exc}.
There are two third-order processes which involve O-O hopping between two $%
p_{\pi }$ orbitals, which lead to a next-NN hopping $V^{\prime }$ between Mo
sites. The band calculations suggest that the equivalent process between Fe
ions is not important and we neglect it. By symmetry, it is clear that the $%
xy$ orbitals cannot hop to NN or next-NN $yz$ or $zx$ orbitals. In addition
(excluding direct hopping involving $d$ orbitals at distances larger than $%
a/2$), the $xy$ orbitals at any site $i$ cannot hop out of the plane,
because its hopping to any O orbital of the sites $i\pm (a/2){\bf \hat{z}}$
is also zero by symmetry. Thus, with a high degree of accuracy, electrons
occupying $xy$ orbitals move only in the $x-y$ plane. Similar considerations
extend to the $yz$ and $zx$ orbitals.

This leads to the following effective Hamiltonian for the movement of
electrons with spin down among $d$ orbitals:

\begin{eqnarray}
H &=&E_{Fe}\sum_{i\alpha }n_{i\alpha }+E_{Mo}\sum_{j\alpha }n_{j\alpha
}+U\sum_{i,\alpha <\beta }n_{i\alpha }n_{i\beta }  \nonumber \\
&&-V\sum_{i\alpha \delta _{\alpha }}(c_{i+\delta _{\alpha }\alpha }^{\dagger
}c_{i\alpha }+{\rm H.c.})-V^{\prime }\sum_{j\alpha \gamma _{\alpha
}}c_{j+\gamma _{\alpha }\alpha }^{\dagger }c_{j\alpha }.  \label{h1}
\end{eqnarray}
Here $c_{i\alpha }^{\dagger }$ creates an electron at the orbital $\alpha
=xy $, $yz$ or $zx$ of site $i$ with spin down, the sum over $i$ ($j$)
extends over the Fe (Mo) sites, $n_{i\alpha }=c_{i\alpha }^{\dagger
}c_{i\alpha }$, and $\delta _{\alpha }$ ($\gamma _{\alpha }$) are vectors
which connect a site with their four NN (next-NN) sites lying in the $\alpha 
$ plane. $U$ is the on-site Coulomb repulsion at the Fe sites. The
corresponding term at the Mo sites is neglected. Eliminating the $d-p_{\pi }$
hopping through a canonical transformation, the values of the other
parameters are:

\begin{eqnarray}
E_{Fe} &=&\epsilon _{Fe}-\frac{4t_{Fe-O}^{2}}{\epsilon _{Fe}-\epsilon _{O}}%
\text{, }E_{Mo}=\epsilon _{Mo}-\frac{4t_{Mo-O}^{2}}{\epsilon _{Mo}-\epsilon
_{O}}\text{, }  \nonumber \\
V &=&\frac{t_{Fe-O}t_{Mo-O}}{2}\left( \frac{1}{\epsilon _{Fe}-\epsilon _{O}}+%
\frac{1}{\epsilon _{Mo}-\epsilon _{O}}\right) \text{,}  \nonumber \\
V^{\prime } &=&\frac{2t_{Mo-O}^{2}t_{O-O}}{(\epsilon _{Mo}-\epsilon _{O})^{2}%
},  \label{par}
\end{eqnarray}
where $t_{O-O}$ is the absolute value of the hopping between the $p_{\pi }$
orbitals of two nearest-neighbor O atoms.

In order to treat the Hamiltonian Eq. (\ref{h1}), we use a simple extension
of the slave-boson theory of Kotliar and Ruckenstein \cite{kot}, to the case
in which there are two sites per elementary cell and three ``colors'' per Fe
site (instead of two representing spin up and down). The Fock space at each
Fe site $i$ is enlarged introducing boson states represented by the creation
operators $e_{i}$ (empty), $s_{i\alpha }$ (singly occupied at orbital $%
\alpha $), $d_{i\alpha \beta }=d_{i\beta \alpha }$ (doubly occupied at
orbitals $\alpha $ and $\beta \neq \alpha $) and $t_{i}$ (triply occupied).
In the combined space $H$ reads:

\begin{eqnarray}
H =E_{Fe}\sum_{i\alpha }n_{i\alpha }+E_{Mo}\sum_{j\alpha }n_{j\alpha
}+U\sum_{i}(\sum_{\alpha <\beta }d_{i\alpha \beta }^{\dagger }d_{i\alpha
\beta }+3t_{i})  \nonumber \\
-V\sum_{i\alpha \delta _{\alpha }}(c_{i+\delta _{\alpha }\alpha }^{\dagger
}c_{i\alpha }z_{i\alpha }+{\rm H.c.})-V^{\prime }\sum_{j\alpha \gamma
_{\alpha }}c_{j+\gamma _{\alpha }\alpha }^{\dagger }c_{j\alpha }  \nonumber
\\
+\sum_{i\alpha }\lambda _{i\alpha }(s_{i\alpha }^{\dagger }s_{i\alpha
}+\sum_{\beta \neq \alpha }d_{i\alpha \beta }^{\dagger }d_{i\alpha \beta
}+t_{i}^{\dagger }t_{i}-n_{i\alpha })  \nonumber \\
+\sum_{i}\lambda _{i}^{\prime }(e_{i}^{\dagger }e_{i}+t_{i}^{\dagger
}t_{i}+\sum_{\alpha }s_{i\alpha }^{\dagger }s_{i\alpha }+\sum_{\alpha <\beta
}d_{i\alpha \beta }^{\dagger }d_{i\alpha \beta }-1),  \label{h2}
\end{eqnarray}
where $z_{i\alpha }^{\dagger }$ represent creation of an electron in the
bosonic part of the Fock space:

\begin{eqnarray}
z_{i\alpha }^{\dagger } =(1-e_{i}^{\dagger }e_{i}-\sum_{\beta \neq \alpha
}s_{i\beta }^{\dagger }s_{i\beta }-d_{i\gamma \eta }^{\dagger }d_{i\gamma
\eta })^{-1/2}    \nonumber \\
\times (s_{i\alpha }^{\dagger }e_{i} 
+\sum_{\beta \neq \alpha
}d_{i\alpha \beta }^{\dagger }s_{i\beta }+t_{i}^{\dagger }d_{i\gamma \eta })
\nonumber \\
\times (1-s_{i\alpha }^{\dagger }s_{i\alpha }-\sum_{\beta \neq \alpha
}d_{i\alpha \beta }^{\dagger }d_{i\alpha \beta }-t_{i}^{\dagger
}t_{i})^{-1/2},  \label{z}
\end{eqnarray}
where $\gamma $ and $\eta $ are the two orbitals different from $\alpha $,
and the last two terms of Eq. (\ref{h2}) were introduced to satisfy the
constraints of vanishing of the corresponding expressions between brackets,
when the energy is minimized with respect to the Lagrange multipliers $%
\lambda _{i\alpha }$ and $\lambda _{i}^{\prime }$. The roots in Eq. (\ref{z}%
) were carefully chosen to satisfy three requirements: i) they are equal to
1 when treated exactly inside this expression, ii) they respect
electron-hole symmetry, and iii) when $U=0$, the saddle point of Eq. (\ref
{h2}) reproduces the exact ground state energy.

In the following we consider the Hamiltonian (\ref{h2}) in the saddle point
approximation, looking at the homogeneous symmetric solution of the
minimization problem, in which all values of the condensed bosons are
independent of site and orbital. The problem reduces to a non interacting
one with renormalized NN hopping $Vz$, where $z$ is the saddle point value
of $z_{i\alpha }^{\dagger }$ (assumed real). We neglect the term in $%
V^{\prime }$. Its effect near the metal-insulator transition is small and
can be taken into account renormalizing $V$. Taking $V^{\prime }=0$ allows
us to obtain analytical expressions for the electronic spectral densities of
states $\rho _{\alpha Fe}$ and $\rho _{\alpha Mo}$ as a function of the
saddle point values of the bosons ($e$, $s$, $d$ and $t$) and multiplier $%
\lambda _{i\alpha }=\lambda $:

\begin{eqnarray}
\rho _{\alpha Fe}(\omega ) =(\omega -E_{Mo})F(\omega )\text{, }\rho
_{\alpha Mo}(\omega )=(\omega -E_{Fe}^{\prime })F(\omega )\text{,}  
\nonumber
\\
\text{with }E_{Fe}^{\prime } =E_{Fe}-\lambda \text{, }F(\omega )=\rho
_{0}(r)/r\text{, }  \nonumber \\
\text{ }r =\text{sgn}(2\omega -E_{Mo}-E_{Fe}^{\prime })\sqrt{(\omega
-E_{Mo})(\omega -E_{Fe}^{\prime })},  
\label{ro}
\end{eqnarray}
and $\rho _{0}(\omega )$ is the spectral density for a square lattice with
NN hopping $Vz$. For simplicity, we replace this density by a constant $\rho
_{0}=1/(2W)$ extending from $-W$ to $W$ with $W=4Vz$. This allows to perform
the integrals analytically and the energy per Fe site takes the form:

\begin{eqnarray}
E(s,d,t) =\frac{E_{Mo}+E_{Fe}}{2}-\frac{3[W^{2}+(E_{Mo}-E_{Fe}^{\prime
})^{2}/4]^{1/2}}{2}  \nonumber \\
+[4W^{2}/9+(E_{Mo}-E_{Fe}^{\prime })^{2}/4]^{1/2} \nonumber \\
+\frac{1}{4}
(1-6m)(E_{Mo}-E_{Fe}-\lambda )  
+3U(d^{2}+t^{2}),  \label{e}
\end{eqnarray}
where $e$ is eliminated using the constraint $1=e^{2}+t^{2}+3(s^{2}+d^{2})$, 
$m=s^{2}+2d^{2}+t^{2}$, and $\lambda $ is eliminated from:

\begin{eqnarray}
\langle n_{i\alpha }\rangle  &=&m=\frac{1}{6}+\frac{E_{Mo}-E_{Fe}^{\prime }}{%
4W}  \nonumber \\
&&\ln \left( \frac{[W^{2}+(E_{Mo}-E_{Fe}^{\prime })^{2}/4]^{1/2}+W}{%
[4W^{2}/9+(E_{Mo}-E_{Fe}^{\prime })^{2}/4]^{1/2}+2W/3}\right) .  \label{lam}
\end{eqnarray}

In Fig. 1 we show the resulting values of $z$ after minimization of $%
E(s,d,t) $, as a function of $U$ for several values of $\Delta
=E_{Mo}-E_{Fe} $. For sufficiently large values of these quantities, $z=0$.
This indicates that the carriers become extremely heavy, the effective band
width is zero, and the system is insulating.

\begin{figure}
\narrowtext
\epsfxsize=3.5truein
\vbox{\hskip 0.05truein \epsffile{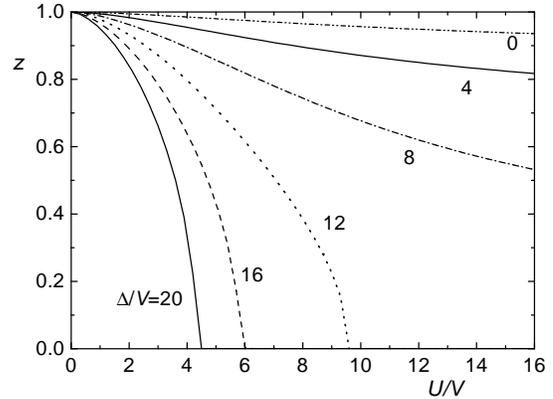}}
\medskip
\caption{Reduction factor of the effective hopping as a function of $U$
for several values of $\Delta =E_{Mo}-E_{Fe}$.}
\end{figure}

For $z\rightarrow 0$ ($W\rightarrow 0$), the expressions (\ref{e}) and (\ref
{lam}) can be expanded and the minimization problem can be further
simplified, allowing a simple analytical description of the metal-insulator
boundary. Differentiating the resulting $E(s,d,t)$ leads to $t=0$ and:

\begin{eqnarray}
\frac{3(E_{Mo}-E_{Fe})}{2\sqrt{38}V} &=&2y+(1+3y^{2})^{1/2}+\frac{1}{%
(1+3y^{2})^{1/2}},  \nonumber \\
\frac{9U}{2\sqrt{38}V} &=&\frac{2}{y}+\frac{3}{(1+3y^{2})^{1/2}},
\label{mit}
\end{eqnarray}
where $y=d/(1-3m)^{1/2}$. As $y$ varies from zero to $\infty $, Eqs. (\ref
{mit}) map the metal-insulator boundary. Due to the fact that $t$ goes more
rapidly to zero than $d$ and $m-1/3$, the problem takes a similar form as
the metal-insulator transition in the cuprates \cite{bal}. Taking the limits
of large or small $y$ in Eqs. (\ref{mit}), asymptotic analytical expressions
for the boundary can be derived:

\begin{eqnarray}
E_{Mo}-E_{Fe} &=&\frac{152(2+\sqrt{3})^{2}}{27}\frac{V^{2}}{U}\text{, }%
U\rightarrow 0\text{;}  \nonumber \\
E_{Mo}-E_{Fe} &=&\frac{4\sqrt{38}}{3}V+\frac{608}{27}\frac{V^{2}}{U}\text{, }%
U\rightarrow \infty  \label{lim}
\end{eqnarray}
One can see from here that the insulating behavior due to Coulomb repulsion
at Fe sites is not possible if $E_{Mo}-E_{Fe}<(4\sqrt{38}/3)V\simeq 8.22V$,
slightly larger than the band width of $\rho _{0}$. This explains the
different behavior of the curves in Fig. 1, depending on the value of $%
(E_{Mo}-E_{Fe})/V$.

\begin{figure}
\narrowtext
\epsfxsize=3.5truein
\vbox{\hskip 0.05truein \epsffile{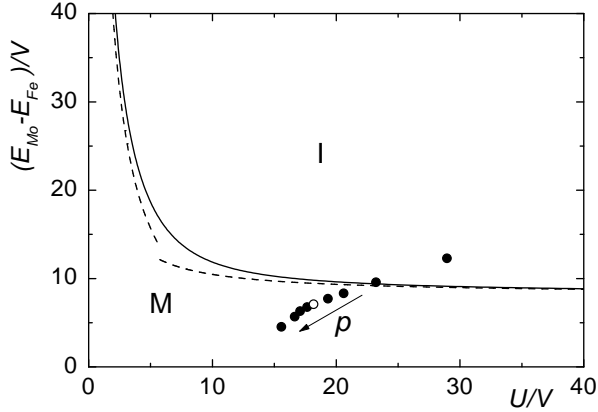}}
\medskip
\caption{Phase diagram of the model in the $E_{Mo}-E_{Fe}$ vs $U$
plane, separating the regions of metallic (M), and insulating (I) behavior.
The dashed lines correspond to the asymptotic analytical expressions (\ref
{lim}) for $U\rightarrow 0$ and $U\rightarrow \infty $. The circles
correspond to the points listed in Table I. Open circle corresponds to the
experimental system at zero pressure. The arrow denotes the direction of
increasing pressure.}
\end{figure}

The resulting metal-insulator phase diagram and the asymptotic expressions (%
\ref{lim}) are represented in Fig. 2. The realization of such a phase
transition depends on the possibility of controlling the parameters of the
Hamiltonian in these materials. As an example, we have studied the effects
of pressure (positive or negative) on the parameters for Sr$_{2}$FeMoO$_{6}$%
. The Madelung potential at Fe sites should be very different of that at Mo
sites due to the large charge difference. In consequence, one expects the
application of pressure or chemical pressure should modify the energy
difference between Fe and Mo $(E_{Mo}-E_{Fe})$ , on one side and also the
hybridization parameter $V$. These parameters were obtained from a fit of
our calculated $t_{2g}$ energy bands for different values of the lattice
parameter, to the bands that result from the Hamiltonian Eq. (\ref{h1}) with 
$V^{\prime }=0$, treated in the Hartree-Fock approximation. In particular
the band that crosses the Fermi energy was adjusted at its bottom (point $%
\Gamma $), and $E_{Mo}-E_{Fe}$ was determined from a fit of the bands at the
L point. The fit for the experimental lattice parameter is shown in Fig. 3.
The bands at higher energies are affected by high energy states not included
in Eq. (\ref{h1}), and a fitting of them is outside our scope. In Table I,
we show the resulting parameters, and the pressure $p$ (obtained from the
numerical derivative of the total energy in the band structure calculation
with respect to the volume) for different lattice parameters $a$. For the
experimental $a=14.91$ \AA\ at $p=0$, we obtain a small but non-zero $p$ due
to the errors of the method. The corresponding points in parameter space are
represented by circles in Fig. 2. One can see that in fact Sr$_{2}$FeMoO$%
_{6} $ is not far from a metal-insulator transition and this can be achieved
applying a negative pressure estimated in -8 GPa.

\begin{figure}
\narrowtext
\epsfxsize=3.5truein
\vbox{\hskip 0.05truein \epsffile{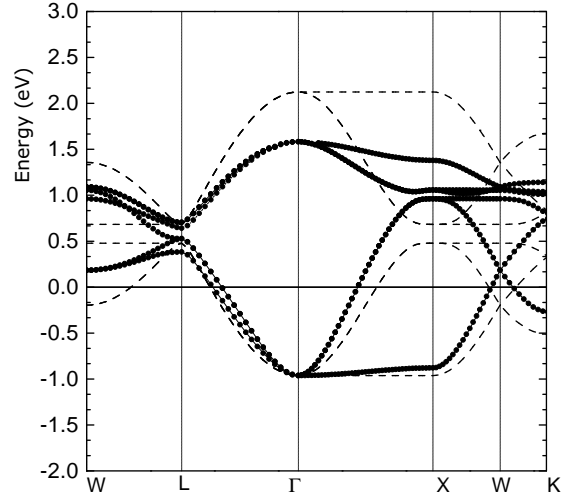}}
\medskip
\caption{Calculated $t_{2g}$ energy bands (full line and solid circles)
and fitting using Eq. (\ref{h1}) with $V^{\prime }=0$ in the Hartree-Fock
approximation (dashed lines). The Fermi energy is at 0 eV. The wave vectors
shown are: W$=(\pi /a,0,2\pi /a)$, L$=(\pi /a,\pi /a,\pi /a)$, $\Gamma
=(0,0,0)$, X$=(0,0,2\pi /a)$, and K$=(3\pi /2a,3\pi /2a,0)$.}
\end{figure}

\begin{tabular}{|c|c|c|c|}
\hline
$a($\AA $)$ & $P($GPa$)$ & $V$(eV) & $E_{Mo}-E_{Fe}$(eV) \\ \hline
{\normalsize 16.51} & {\normalsize -12.47} & {\normalsize 0.2417} & 
{\normalsize 2.967} \\ \hline
{\normalsize 15.71} & {\normalsize -8.37} & {\normalsize 0.3015} & 
{\normalsize 2.886} \\ \hline
{\normalsize 15.31} & {\normalsize -4.57} & {\normalsize 0.3396} & 
{\normalsize 2.824} \\ \hline
{\normalsize 15.11} & {\normalsize -1.69} & {\normalsize 0.3618} & 
{\normalsize 2.789 } \\ \hline
{\normalsize 14.91} & {\normalsize 1.72} & {\normalsize 0.3850} & 
{\normalsize 2.727 } \\ \hline
{\normalsize 14.81} & {\normalsize 3.52} & {\normalsize 0.3964} & 
{\normalsize 2.681 } \\ \hline
{\normalsize 14.71} & {\normalsize 5.36} & {\normalsize 0.4096} & 
{\normalsize 2.577} \\ \hline
{\normalsize 14.61} & {\normalsize 7.17} & {\normalsize 0.4203} & 
{\normalsize 2.385} \\ \hline
{\normalsize 14.51} & {\normalsize 9.07} & {\normalsize 0.4494} & 
{\normalsize 2.029} \\ \hline
\end{tabular}

Table I. Pressure and parameters of Eq. (\ref{h1}) for different lattice parameters.

\vskip2pc

In summary, combining band structure calculations, with a slave boson
technique (equivalent to the Gutzwiller approximation) \cite{kot} to treat
strong correlations, we have studied the possibility of a metal-insulator
transition in Sr$_{2}$FeMoO$_{6}$.We expect that this approach can be used
to understand the metallic or insulating behavior in other double
perovskites.

We thank G. Zampieri for useful discussions. We are partially supported by
CONICET. This work was sponsored by PICT 03-06343 of ANPCyT and PIP 4952/96
of CONICET.


\begin{references}
\bibitem{Kobayashi}  K.I. Kobayashi, T. Kimura, H. Sawada, K. Terakura and
Y. Tokura, Nature {\bf 395}, 677 (1998).

\bibitem{Boc}  A.E. Bocquet, T. Mizokawa, T. Saitoh, H. Namatame, and A.
Fujimori, Phys. Rev. B {\bf 46}, 3771 (1992).

\bibitem{Millis}  A. Chattopadhyay and A. J. Millis, cond-mat/0006208.

\bibitem{Guille}  M.S. Moreno, J.E. Gayone, A. Caneiro, D. Niebieskiwiat,
R.D. Sanchez, and G. Zampieri, submitted to Solid State Commun.

\bibitem{exc}  M.E. Simon, A.A. Aligia, C.D. Batista, E. Gagliano, and F.
Lema, Phys. Rev. B {\bf 54}, R3780 (1996).

\bibitem{gar}  B. Garc\'{i}a-Landa, C. Ritter, M.R. Ibarra, J. Blasco, P.A.
Algarabel, R. Mahendiran, and J. Garc\'{i}a, Solid State Commun. 110, 435
(1999)

\bibitem{Gopa}  J. Gopalakrishnan, A. Chattopadhyay, S.B. Ogale, T.
Venkatesan, R.L. Greene, A.J. Millis, K. Ramesha, B. Hannoyer, and G.
Marest, cond-mat/0004315.

\bibitem{Itoh}  M. Itoh, I. Ohta, and Y. Inaguma, Mat. Science Ing. B {\bf 41%
}, 55 (1996).

\bibitem{kot}  G. Kotliar and A.E. Ruckenstein, Phys. Rev. Lett. {\bf 57},
1362 (1986).

\bibitem{bal}  C.A. Balseiro, M. Avignon, A.G. Rojo, and B. Alascio, Phys.
Rev. Lett. {\bf 62}, 2624 (1989); {\bf 63}, 696 (E) (1989).

\bibitem{blaha95}  P.\ Blaha, K.\ Schwarz, and J.\ Luitz, WIEN97, Vienna
University of Technology, 1997. (Improved and updated Unix version of the
original copyrighted WIEN-code, which was published by P.\ Blaha, K.\
Schwarz, P.\ Soratin, and S.\ B.\ Trickey, in Comput.\ Phys.\ Commun.\ {\bf %
59}, 399 (1990)).

\bibitem{singhbook}  D.\ Singh, {\it Plane Waves, Pseudopotentials, and the
LAPW Method} (Kluwer Academic, New York, 1994).

\bibitem{perdew96}  J.\ P.\ Perdew, S.\ Burke and M.\ Ernzerhof, Phys.\
Rev.\ Lett.\ {\bf 77}, 3865 (1996).

\bibitem{singh91}  D.\ Singh, Phys.\ Rev.\ B {\bf 43}, 6388 (1991).

\bibitem{sim2}  M.E. Simon, A.A. Aligia, and E. Gagliano, Phys. Rev. B {\bf %
56}, 5637 (1997); references therein. H. Rosner, H. Eschrig, R. Hayn, S.-L.
Drechsler, and J. M\'{a}lek, {\it ibid }{\bf 56}, 3402 (1997); references
therein.
\end{references}
\end{document}